\magnification= 1200

\input hyperbasics
\catcode`\@=11
\def\unredoffs{\voffset=13mm \hoffset=6.5truemm} 
\def\redoffs{\voffset=-12.truemm\hoffset=-3truemm} 

\newif\ifbookmode
\bookmodefalse
%
\newwrite\inx
\def\book{\bookmodetrue\immediate\openout\inx=\jobname.inx%
 \hsize=120mm\vsize=195mm} 
\def\@#1{\noindent\ifbookmode\write\inx{#1,\space
\number\pageno.\par}\fi}

\newbox\leftpage \newdimen\fullhsize \newdimen\hstitle \newdimen\hsbody
\newdimen\hdim
\hfuzz=1pt
\ifx\hyperdef\UNd@FiNeD\def\hyperdef#1#2#3#4{#4}\def\hyperref#1#2#3#4{#4}\fi
\def\bigans{b }
\def\answ{b }
\ifx\answ\bigans\message{(Format simple colonne 12pts.}
\magnification=1200 \unredoffs
\hsize=122.5mm\vsize=182.5mm
\hsbody=\hsize \hstitle=\hsize 
\else\message{(Format simple colonne, 10pts.} \let\l@r=L
\magnification=1000 
\redoffs%
\hsize=122.5mm\vsize=182.5mm
\hsbody=\hsize \hstitle=\hsize 
\fi
%

\def\sla#1{\mkern-1.5mu\raise0.4pt\hbox{$\not$}\mkern1.2mu #1\mkern 0.7mu}
\def\Dbar{\mkern-1.5mu\raise0.4pt\hbox{$\not$}\mkern-.1mu {\rm D}\mkern.1mu}
\def\Abar{\mkern1.mu\raise0.4pt\hbox{$\not$}\mkern-1.3mu A\mkern.1mu}
\def\Bbar{\mkern-0.mu\raise0.4pt\hbox{$\not$}\mkern-.3mu B\mkern 0.6mu}
\newskip\tableskipamount \tableskipamount=8pt plus 3pt minus 3pt

\newdimen\chapskip
\chapskip=17.5mm
\font\twbfx=cmbx10 scaled 1200
\font\chapfnt=cmbx10 scaled 1440
\font\chapbxten=cmbx10
\font\chapbxseven=cmbx7

\font\ssbx=cmssbx10  

\font\chaprm=cmr10 scaled 1440
\font\chaprmscript=cmr10
\font\chaprmseven=cmr7
\font\chapssfnt=cmssbx10 scaled 1440

\font\chapibfnt=cmmib10 scaled 1440
\font\chapmifnt=cmmi10 scaled 1440
\font\chapsyfnt=cmsy10 scaled 1440
\font\chapexfnt=cmex10 scaled 1440
\font\chapibten=cmmib10
\font\chapibseven=cmmib7
\font\chapmiscript=cmmi10  
\font\chapsyscript=cmsy10  
\font\chapexscript=cmex10
\font\chapmiseven=cmmi7
\font\chapsyseven=cmsy7
\font\chapexseven=cmex7  
\def\chapfont{
\textfont0=\chaprm\scriptfont0=\chaprmscript\scriptscriptfont0=\chaprmseven
\textfont1=\chapmifnt \scriptfont1=\chapmiscript  \scriptscriptfont1=\chapmiseven
\textfont2=\chapsyfnt \scriptfont2=\chapsyscript\scriptscriptfont2=\chapsyseven
\textfont3=\chapexfnt \scriptfont3=\chapexscript \scriptscriptfont3=\chapexseven
\textfont\bffam=\chapfnt \scriptfont\bffam=\chapbxten
\scriptscriptfont\bffam=\chapbxseven
\def\bf{\fam\bffam\chapfnt} 
\textfont4=\chapibfnt \scriptfont4=\chapibten \scriptscriptfont4=\chapibseven
\abovedisplayskip=17pt plus 5pt minus 13pt
\belowdisplayskip=\abovedisplayskip
\normalbaselineskip=17pt
\setbox\strutbox=\hbox{\vrule height12.2pt depth5.0pt width0pt}
\normalbaselines \chapssfnt}

\font\caprm=cmr9
\font\capit=cmti9
\font\capbf=cmbx9
\font\capsl=cmsl9
\font\capmi=cmmi9
\font\capex=cmex9
\font\capsy=cmsy9

\def\makeheadline{\vbox to 0pt{\vskip-22.5pt
\line{\vbox to8.5pt{}\the\headline}\vss}\nointerlineskip}
\font\headrm=cmr10

\font\sixrm=cmr6
\font\fiverm=cmr5
\font\sixmi=cmmi6
\font\fivemi=cmmi5
\font\sixsy=cmsy6
\font\fivesy=cmsy5
\font\sixbf=cmbx6
\font\fivebf=cmbx5
\skewchar\capmi='177 \skewchar\sixmi='177 \skewchar\fivemi='177
\skewchar\capsy='60 \skewchar\sixsy='60 \skewchar\fivesy='60


%
%

\catcode`\@=11
\font\tenbi=cmmib10 
\font\ninebi=cmmib9
\font\sevenbi=cmmib7 
\font\fivebi=cmmib5
\textfont4=\tenbi \scriptfont4=\sevenbi \scriptscriptfont4=\fivebi
\newfam\mibfam
\textfont\mibfam=\tenbi \scriptfont\mibfam=\sevenbi \scriptscriptfont\mibfam=\fivebi
\def\Bfg{\ifmmode\let\next\Bfg@\else
 \def\next{\errmessage{Use \string\Bfg\space only in math mode}}\fi\next}
\def\Bfg@#1{{\Bfg@@{#1}}}
\def\Bfg@@#1{\fam\mibfam#1}
\skewchar\tenbi='177\skewchar\sevenbi='177
\mathchardef\alpha="710B
\mathchardef\beta="710C 
\mathchardef\gamma="710D
\mathchardef\delta="710E
\mathchardef\epsilon="710F  
\mathchardef\zeta="7110  
\mathchardef\eta="7111 
\mathchardef\theta="7112 
\mathchardef\iota="7113 
\mathchardef\kappa="7114 
\mathchardef\lambda="7115 
\mathchardef\mu="7116 
\mathchardef\nu="7117 
\mathchardef\xi="7118 
\mathchardef\pi="7119 
\mathchardef\rho="711A  
\mathchardef\sigma="711B 
\mathchardef\tau="711C 
\mathchardef\upsilon="711D 
\mathchardef\phi="711E 
\mathchardef\chi="711F 
\mathchardef\psi="7120 
\mathchardef\omega="7121 
\mathchardef\varepsilon="7122 
\mathchardef\vartheta="7123
\mathchardef\varpi="7124 
\mathchardef\varrho="7125 
\mathchardef\varsigma="7126 
\mathchardef\varphi="7127
\mathchardef\alphab="040B
\mathchardef\betab="040C 
\mathchardef\gammab="040D
\mathchardef\deltab="040E
\mathchardef\epsilonb="040F  
\mathchardef\zetab="0410  
\mathchardef\etab="0411 
\mathchardef\thetab="0412 
\mathchardef\iotab="0413 
\mathchardef\kappab="0414 
\mathchardef\lambdab="0415 
\mathchardef\mub="0416 
\mathchardef\nub="0417 
\mathchardef\xib="0418 
\mathchardef\pib="0419 
\mathchardef\rhob="041A  
\mathchardef\sigmab="041B 
\mathchardef\taub="041C 
\mathchardef\upsilonb="041D 
\mathchardef\phib="041E 
\mathchardef\chib="041F 
\mathchardef\psib="0420 
\mathchardef\omegab="0421 
\mathchardef\varepsilonb="0422 
\mathchardef\varthetab="0423
\mathchardef\varpib="0424 
\mathchardef\varrhob="0425 
\mathchardef\varsigmab="0426 
\mathchardef\varphib="0427 
\font\tenmsa=msam10
\font\sevenmsa=msam7
\font\fivemsa=msam5
\font\tenmsb=msbm10
\font\sevenmsb=msbm7
\font\fivemsb=msbm5
\newfam\msafam
\newfam\msbfam
\textfont\msafam=\tenmsa  \scriptfont\msafam=\sevenmsa
  \scriptscriptfont\msafam=\fivemsa
\textfont\msbfam=\tenmsb  \scriptfont\msbfam=\sevenmsb
  \scriptscriptfont\msbfam=\fivemsb

\def\hexnumber@#1{\ifcase#1 0\or1\or2\or3\or4\or5\or6\or7\or8\or9\or
	A\or B\or C\or D\or E\or F\fi }
\font\teneuf=eufm10
\font\seveneuf=eufm7
\font\fiveeuf=eufm5
\newfam\euffam
\textfont\euffam=\teneuf
\scriptfont\euffam=\seveneuf
\scriptscriptfont\euffam=\fiveeuf
\def\frak{\ifmmode\let\next\frak@\else
 \def\next{\Err@{Use \string\frak\space only in math mode}}\fi\next}
\def\goth{\ifmmode\let\next\frak@\else
 \def\next{\Err@{Use \string\goth\space only in math mode}}\fi\next}
\def\frak@#1{{\frak@@{#1}}}
\def\frak@@#1{\fam\euffam#1}

\edef\msa@{\hexnumber@\msafam}
\edef\msb@{\hexnumber@\msbfam}

\mathchardef\boxdot="2\msa@00
\mathchardef\boxplus="2\msa@01
\mathchardef\boxtimes="2\msa@02
\mathchardef\square="0\msa@03
\mathchardef\blacksquare="0\msa@04
\mathchardef\centerdot="2\msa@05
\mathchardef\lozenge="0\msa@06
\mathchardef\blacklozenge="0\msa@07
\mathchardef\circlearrowright="3\msa@08
\mathchardef\circlearrowleft="3\msa@09
\mathchardef\rightleftharpoons="3\msa@0A
\mathchardef\leftrightharpoons="3\msa@0B
\mathchardef\boxminus="2\msa@0C
\mathchardef\Vdash="3\msa@0D
\mathchardef\Vvdash="3\msa@0E
\mathchardef\vDash="3\msa@0F
\mathchardef\twoheadrightarrow="3\msa@10
\mathchardef\twoheadleftarrow="3\msa@11
\mathchardef\leftleftarrows="3\msa@12
\mathchardef\rightrightarrows="3\msa@13
\mathchardef\upuparrows="3\msa@14
\mathchardef\downdownarrows="3\msa@15
\mathchardef\upharpoonright="3\msa@16

\mathchardef\downharpoonright="3\msa@17
\mathchardef\upharpoonleft="3\msa@18
\mathchardef\downharpoonleft="3\msa@19
\mathchardef\rightarrowtail="3\msa@1A
\mathchardef\leftarrowtail="3\msa@1B
\mathchardef\leftrightarrows="3\msa@1C
\mathchardef\rightleftarrows="3\msa@1D
\mathchardef\Lsh="3\msa@1E
\mathchardef\Rsh="3\msa@1F
\mathchardef\rightsquigarrow="3\msa@20
\mathchardef\leftrightsquigarrow="3\msa@21
\mathchardef\looparrowleft="3\msa@22
\mathchardef\looparrowright="3\msa@23
\mathchardef\circeq="3\msa@24
\mathchardef\succsim="3\msa@25
\mathchardef\gtrsim="3\msa@26
\mathchardef\gtrapprox="3\msa@27
\mathchardef\multimap="3\msa@28
\mathchardef\therefore="3\msa@29
\mathchardef\because="3\msa@2A
\mathchardef\doteqdot="3\msa@2B

\mathchardef\triangleq="3\msa@2C
\mathchardef\precsim="3\msa@2D
\mathchardef\lesssim="3\msa@2E
\mathchardef\lessapprox="3\msa@2F
\mathchardef\eqslantless="3\msa@30
\mathchardef\eqslantgtr="3\msa@31
\mathchardef\curlyeqprec="3\msa@32
\mathchardef\curlyeqsucc="3\msa@33
\mathchardef\preccurlyeq="3\msa@34
\mathchardef\leqq="3\msa@35
\mathchardef\leqslant="3\msa@36
\mathchardef\lessgtr="3\msa@37
\mathchardef\backprime="0\msa@38
\mathchardef\risingdotseq="3\msa@3A
\mathchardef\fallingdotseq="3\msa@3B
\mathchardef\succcurlyeq="3\msa@3C
\mathchardef\geqq="3\msa@3D
\mathchardef\geqslant="3\msa@3E
\mathchardef\gtrless="3\msa@3F
\mathchardef\sqsubset="3\msa@40
\mathchardef\sqsupset="3\msa@41
\mathchardef\vartriangleright="3\msa@42
\mathchardef\vartriangleleft="3\msa@43
\mathchardef\trianglerighteq="3\msa@44
\mathchardef\trianglelefteq="3\msa@45
\mathchardef\bigstar="0\msa@46
\mathchardef\between="3\msa@47
\mathchardef\blacktriangledown="0\msa@48
\mathchardef\blacktriangleright="3\msa@49
\mathchardef\blacktriangleleft="3\msa@4A
\mathchardef\vartriangle="0\msa@4D
\mathchardef\blacktriangle="0\msa@4E
\mathchardef\triangledown="0\msa@4F
\mathchardef\eqcirc="3\msa@50
\mathchardef\lesseqgtr="3\msa@51
\mathchardef\gtreqless="3\msa@52
\mathchardef\lesseqqgtr="3\msa@53
\mathchardef\gtreqqless="3\msa@54
\mathchardef\Rrightarrow="3\msa@56
\mathchardef\Lleftarrow="3\msa@57
\mathchardef\veebar="2\msa@59
\mathchardef\barwedge="2\msa@5A
\mathchardef\doublebarwedge="2\msa@5B
\mathchardef\angle="0\msa@5C
\mathchardef\measuredangle="0\msa@5D
\mathchardef\sphericalangle="0\msa@5E
\mathchardef\varpropto="3\msa@5F
\mathchardef\smallsmile="3\msa@60
\mathchardef\smallfrown="3\msa@61
\mathchardef\Subset="3\msa@62
\mathchardef\Supset="3\msa@63
\mathchardef\Cup="2\msa@64

\mathchardef\Cap="2\msa@65

\mathchardef\curlywedge="2\msa@66
\mathchardef\curlyvee="2\msa@67
\mathchardef\leftthreetimes="2\msa@68
\mathchardef\rightthreetimes="2\msa@69
\mathchardef\subseteqq="3\msa@6A
\mathchardef\supseteqq="3\msa@6B
\mathchardef\bumpeq="3\msa@6C
\mathchardef\Bumpeq="3\msa@6D
\mathchardef\lll="3\msa@6E

\mathchardef\ggg="3\msa@6F

\mathchardef\circledS="0\msa@73
\mathchardef\pitchfork="3\msa@74
\mathchardef\dotplus="2\msa@75
\mathchardef\backsim="3\msa@76
\mathchardef\backsimeq="3\msa@77
\mathchardef\complement="0\msa@7B
\mathchardef\intercal="2\msa@7C
\mathchardef\circledcirc="2\msa@7D
\mathchardef\circledast="2\msa@7E
\mathchardef\circleddash="2\msa@7F
\def\ulcorner{\delimiter"4\msa@70\msa@70 }
\def\urcorner{\delimiter"5\msa@71\msa@71 }
\def\llcorner{\delimiter"4\msa@78\msa@78 }
\def\lrcorner{\delimiter"5\msa@79\msa@79 }
\def\yen{\mathhexbox\msa@55 }
\def\checkmark{\mathhexbox\msa@58 }
\def\circledR{\mathhexbox\msa@72 }
\def\maltese{\mathhexbox\msa@7A }
\mathchardef\lvertneqq="3\msb@00
\mathchardef\gvertneqq="3\msb@01
\mathchardef\nleq="3\msb@02
\mathchardef\ngeq="3\msb@03
\mathchardef\nless="3\msb@04
\mathchardef\ngtr="3\msb@05
\mathchardef\nprec="3\msb@06
\mathchardef\nsucc="3\msb@07
\mathchardef\lneqq="3\msb@08
\mathchardef\gneqq="3\msb@09
\mathchardef\nleqslant="3\msb@0A
\mathchardef\ngeqslant="3\msb@0B
\mathchardef\lneq="3\msb@0C
\mathchardef\gneq="3\msb@0D
\mathchardef\npreceq="3\msb@0E
\mathchardef\nsucceq="3\msb@0F
\mathchardef\precnsim="3\msb@10
\mathchardef\succnsim="3\msb@11
\mathchardef\lnsim="3\msb@12
\mathchardef\gnsim="3\msb@13
\mathchardef\nleqq="3\msb@14
\mathchardef\ngeqq="3\msb@15
\mathchardef\precneqq="3\msb@16
\mathchardef\succneqq="3\msb@17
\mathchardef\precnapprox="3\msb@18
\mathchardef\succnapprox="3\msb@19
\mathchardef\lnapprox="3\msb@1A
\mathchardef\gnapprox="3\msb@1B
\mathchardef\nsim="3\msb@1C
\mathchardef\ncong="3\msb@1D

\mathchardef\varsubsetneq="3\msb@20
\mathchardef\varsupsetneq="3\msb@21
\mathchardef\nsubseteqq="3\msb@22
\mathchardef\nsupseteqq="3\msb@23
\mathchardef\subsetneqq="3\msb@24
\mathchardef\supsetneqq="3\msb@25
\mathchardef\varsubsetneqq="3\msb@26
\mathchardef\varsupsetneqq="3\msb@27
\mathchardef\subsetneq="3\msb@28
\mathchardef\supsetneq="3\msb@29
\mathchardef\nsubseteq="3\msb@2A
\mathchardef\nsupseteq="3\msb@2B
\mathchardef\nparallel="3\msb@2C
\mathchardef\nmid="3\msb@2D
\mathchardef\nshortmid="3\msb@2E
\mathchardef\nshortparallel="3\msb@2F
\mathchardef\nvdash="3\msb@30
\mathchardef\nVdash="3\msb@31
\mathchardef\nvDash="3\msb@32
\mathchardef\nVDash="3\msb@33
\mathchardef\ntrianglerighteq="3\msb@34
\mathchardef\ntrianglelefteq="3\msb@35
\mathchardef\ntriangleleft="3\msb@36
\mathchardef\ntriangleright="3\msb@37
\mathchardef\nleftarrow="3\msb@38
\mathchardef\nrightarrow="3\msb@39
\mathchardef\nLeftarrow="3\msb@3A
\mathchardef\nRightarrow="3\msb@3B
\mathchardef\nLeftrightarrow="3\msb@3C
\mathchardef\nleftrightarrow="3\msb@3D
\mathchardef\divideontimes="2\msb@3E
\mathchardef\varnothing="0\msb@3F
\mathchardef\nexists="0\msb@40
\mathchardef\mho="0\msb@66
\mathchardef\eth="0\msb@67
\mathchardef\eqsim="3\msb@68
\mathchardef\beth="0\msb@69
\mathchardef\gimel="0\msb@6A
\mathchardef\daleth="0\msb@6B
\mathchardef\lessdot="3\msb@6C
\mathchardef\gtrdot="3\msb@6D
\mathchardef\ltimes="2\msb@6E
\mathchardef\rtimes="2\msb@6F
\mathchardef\shortmid="3\msb@70
\mathchardef\shortparallel="3\msb@71
\mathchardef\smallsetminus="2\msb@72
\mathchardef\thicksim="3\msb@73
\mathchardef\thickapprox="3\msb@74
\mathchardef\approxeq="3\msb@75
\mathchardef\succapprox="3\msb@76
\mathchardef\precapprox="3\msb@77
\mathchardef\curvearrowleft="3\msb@78
\mathchardef\curvearrowright="3\msb@79
\mathchardef\digamma="0\msb@7A
\mathchardef\varkappa="0\msb@7B
\mathchardef\hslash="0\msb@7D
\mathchardef\hbar="0\msb@7E
\mathchardef\backepsilon="3\msb@7F
\def\Bbb{\ifmmode\let\next\Bbb@\else
 \def\next{\errmessage{Use \string\Bbb\space only in math mode}}\fi\next}
\def\Bbb@#1{{\Bbb@@{#1}}}
\def\Bbb@@#1{\fam\msbfam#1}

\def\elevenpoint{
\textfont0=\caprm \scriptfont0=\sixrm \scriptscriptfont0=\fiverm
\def\rm{\fam0\caprm}
\textfont1=\capmi \scriptfont1=\sixmi \scriptscriptfont1=\fivemi
\textfont2=\capsy \scriptfont2=\sixsy \scriptscriptfont2=\fivesy
\textfont3=\capex \scriptfont3=\capex \scriptscriptfont3=\capex
\textfont\itfam=\capit \def\it{\fam\itfam\capit} 
\textfont\slfam=\capsl  \def\sl{\fam\slfam\capsl} 
\textfont\bffam=\capbf \scriptfont\bffam=\sixbf
\scriptscriptfont\bffam=\fivebf
\def\bf{\fam\bffam\capbf} 
\textfont4=\ninebi \scriptfont4=\sevenbi \scriptscriptfont4=\fivebi
\abovedisplayskip=11pt plus 3pt minus 8pt
\belowdisplayskip=\abovedisplayskip
\smallskipamount=2.7pt plus 1pt minus 1pt
\medskipamount=5.4pt plus 2pt minus 2pt
\bigskipamount=10.8pt plus 3.6pt minus 3.6pt
\normalbaselineskip=11pt
\setbox\strutbox=\hbox{\vrule height7.8pt depth3.2pt width0pt}
\normalbaselines \rm}
\catcode`\@=12
\def\sla#1{\mkern-1.5mu\raise0.4pt\hbox{$\not$}\mkern1.2mu #1\mkern 0.7mu}
\def\Dbar{\mkern-1.5mu\raise0.4pt\hbox{$\not$}\mkern-.1mu {\rm D}\mkern.1mu}
\def\Abar{\mkern1.mu\raise0.4pt\hbox{$\not$}\mkern-1.3mu A\mkern.1mu}
\nopagenumbers
\def\makeheadline{\vbox to 0pt{\vskip-27pt
\line{\vbox to8.5pt{}\the\headline}\vss}\nointerlineskip}
\def\subsectionname{}
\def\sectionname{}
\headline={\ifnum\pageno=1\hfill\else\draftdate\hfil{\headrm\folio}%
\hfil\hphantom{\draftdate}\fi}	 


\newcount\yearltd\yearltd=\year\advance\yearltd by -2000
\newif\ifdraftmode
\draftmodefalse
\def\draft{\draftmodetrue{\count255=\time\divide\count255 by 60
\xdef\hourmin{\number\count255} 
  \multiply\count255 by-60\advance\count255 by\time
  \xdef\hourmin{\hourmin:\ifnum\count255<10 0\fi\the\count255}}}
\def\draftdate{\ifdraftmode{\headrm\quad (\jobname,\ le
\number\day/\number\month/\number\yearltd\ \ \hourmin)}\else{}\fi} 
\newif\iffrancmode
\francmodefalse

\def\d{{\rm d}}

\chardef\sigmat=27

\def\frac#1#2{{\textstyle{#1\over#2}}}

\def\leaderfill{\leaders\hbox to 1em{\hss.\hss}\hfill}
\catcode`\@=11

\def\deqalignno#1{\displ@y\tabskip\centering \halign to
\displaywidth{\hfil$\displaystyle{##}$\tabskip0pt&$\displaystyle{{}##}$
\hfil\tabskip0pt &\quad
\hfil$\displaystyle{##}$\tabskip0pt&$\displaystyle{{}##}$ 
\hfil\tabskip\centering& \llap{$##$}\tabskip0pt \crcr #1 \crcr}}
\def\deqalign#1{\null\,\vcenter{\openup\jot\m@th\ialign{
\strut\hfil$\displaystyle{##}$&$\displaystyle{{}##}$\hfil
&&\quad\strut\hfil$\displaystyle{##}$&$\displaystyle{{}##}$
\hfil\crcr#1\crcr}}\,}
\def\xlabel#1{\expandafter\xl@bel#1}\def\xl@bel#1{#1}
\def\label#1{\l@bel #1{\hbox{}}}
\def\l@bel#1{\ifx\UNd@FiNeD#1\message{label \string#1 is undefined.}%
\xdef#1{?.? }\fi{\let\hyperref=\relax\xdef\next{#1}}%
\ifx\next\em@rk\def\next{}%
\else\def\next{#1}\fi\next}
\def\DefWarn#1{\ifx\UNd@FiNeD#1\else
\immediate\write16{*** WARNING: the label \string#1 is already defined%
***}\fi}%
\newread\ch@ckfile
\def\cinput#1{\def\filen@me{#1 }
\immediate\openin\ch@ckfile=\filen@me
\ifeof\ch@ckfile\message{<< (\filen@me\ DOES NOT EXIST in this pass)>>}\else%
\closein \ch@ckfile\input\filen@me\fi}
\ifx\UNd@FiNeD\prefix\def\prefix{}\fi 
\newread\ch@ckfile
\immediate\openin\ch@ckfile=\jobname.def
\ifeof\ch@ckfile\message{<< (\jobname.def DOES NOT EXIST in this pass) >>}
\else
\def\DefWarn#1{}%
\closein \ch@ckfile%
\input\jobname.def\fi
\def\listcontent{
\immediate\openin\ch@ckfile=\jobname.tab 
\ifeof\ch@ckfile\message{no file \jobname.tab, no table of contents this
pass}%
\else\closein\ch@ckfile%
\def\sectionname{\iffrancmode Table des
mati\`eres \else Contents\fi}
\centerline{\twbfx\iffrancmode Table des
mati\`eres \else Contents\fi}\nobreak\medskip%
{\baselineskip=12pt\parskip=0pt\catcode`\@=11\input\jobname.tab
\catcode`\@=12\bigbreak\bigskip}\fi}
\newcount\nochapter
\newcount\nosection
\newcount\nosubsection
\newcount\neqno
\newcount\notenumber
\newcount\nofigure
\newcount\notable
\newcount\noexerc
\newcount\fpage
\newcount\firstpage
\newif\ifappmode
\newwrite\equa

\newdimen\hulp
\def\maketitle#1{
\edef\oneliner##1{\centerline{##1}}
\edef\twoliner##1{\vbox{\parindent=0pt\leftskip=0pt plus 1fill\rightskip=0pt
plus 1fill 
                     \parfillskip=0pt\relax##1}} 
\setbox0=\vbox{#1}\hulp=0.5\hsize
                 \ifdim\wd0<\hulp\oneliner{#1}\else
                 \twoliner{#1}\fi}
\def\preprint#1{\ifdraftmode\gdef\prepname{\jobname/#1}\else%
\gdef\prepname{#1}\fi\hfill{
\expandafter{\prepname}}\vskip20mm} 
\def\title#1\par{\gdef\titlename{#1}
\global\firstpage=\pageno
\nosection=0
\mark{\the\nosection}
\maketitle{\ssbx\uppercase\expandafter{\titlename}}
\vskip17mm
\nochapter=0
\neqno=0
\notenumber=0
\nofigure=0
\notable=0
\def\prefix{}
\appmodefalse
\mark{\the\nochapter}
\message{#1}
\immediate\openout\equa=\jobname.equ %
}
\ifbookmode%
\headline={\ifnum\pageno=\firstpage\hfill\else\ifodd\pageno\rightheadline
\else\leftheadline\fi\fi}
\else
\headline={\ifnum\pageno=\firstpage\hfill\else\draftdate\hfil{\headrm\folio}%
\hfil\hphantom{\draftdate}\fi}\fi 
\def\abstract{\vskip8mm\iffrancmode\centerline{R\'ESUM\'E}\else%
\centerline{ABSTRACT}\fi \smallskip \begingroup\narrower
\elevenpoint\baselineskip10pt} 

\def\section#1\par{\vskip0pt plus.1\vsize\penalty-100\vskip0pt plus-.1
\vsize\bigskip\vglue\parskip\par%
\global\fpage=\pageno
\ifnum\nochapter=0\ifappmode\relax\else\writetoc
\fi\fi
\advance\nochapter by 1\global\nosection=0\global\neqno=0
\gdef\sectionname{#1}
\vbox{\noindent\twbfx{\hyperdef\hypernoname{section}{\prefix\the\nochapter}%
{\prefix\the\nochapter}\ #1}}%
\message{\prefix\the\nochapter\ \sectionname}%
\writetoca{{\string\hyperref{}{section}{\prefix\the\nochapter}%
{\prefix\the\nochapter}} {#1}}%
\bigskip\noindent%
}

\def\appendix#1#2\par{\bigbreak\nochapter=0
\appmodetrue
\notenumber=0
\neqno=0
\def\prefix{A}
\mark{\the\nochapter}
\message{APPENDICES}
{\hyperdef\hypernoname{section}{\prefix}{ 
\leftline{\uppercase\expandafter{#1}}
\leftline{\uppercase\expandafter{#2}}}}
\noindent\nonfrenchspacing
\writetoca{\string\hyperref{}{section}{\prefix}{Appendices}.\ #1 \ #2}%
}
\def\subsection#1\par{\vskip0pt plus.05\vsize\penalty-100\vskip0pt
plus-.05\vsize\bigskip\vskip1mm\advance\nosection by 1
\global\nosubsection=0
\def\subsectionname{#1}
\mark{#1}
\message{\the\nochapter.\the\nosection\ #1}
\vbox{\noindent\bf{\hyperdef\hypernoname{section}{\prefix\the\nochapter.%
\the\nosection}{\prefix\the\nochapter.\the\nosection\ #1}}}\vskip1mm%
\smallskip\noindent%
\writetoca{{\string\hyperref{}{section}{\prefix\the\nochapter.%
\the\nosection}{\prefix\the\nochapter.\the\nosection}} {#1}}%
} 
\def\ssubsection#1\par{\vskip0pt plus.05\vsize\penalty-100\vskip0pt%
plus-.05\vsize\medskip\vskip\parskip\advance\nosubsection by 1%
\message{\the\nochapter.\the\nosection.\the\nosubsection\ #1}%
\vbox{\noindent\bf{\hyperdef\hypernoname{section}{\prefix\the\nochapter.%
\the\nosection.\the\nosubsection}{\prefix\the\nochapter.\the\nosection.%
\the\nosubsection\ \bf #1}}}%
\smallskip\noindent%
\writetoca{{\string\hyperref{}{section}{\prefix\the\nochapter.%
\the\nosection.\the\nosubsection}{\prefix\the\nochapter.\the\nosection.%
\the\nosubsection}} {#1}}%
}   
%
\def\note #1{\advance\notenumber by 1
\footnote{$^{\the\notenumber}$}{\sevenrm #1}} 

\parindent=1em 
\newinsert\margin
\dimen\margin=\maxdimen
\count\margin=0 \skip\margin=0pt
\def\sslbl#1{\DefWarn#1%
\ifdraftmode{\leavevmode\vadjust{\smash%
{\line{{\escapechar=` \hfill\rlap{\sevenrm\hskip1mm\string#1}}}}}}%
\fi 
\ifnum\nochapter=0%
\if\prefix{}\xdef#1{}%
\edef\ewrite{\write\equa{{\string#1}}%
\write\equa{}}\ewrite%
\else
\xdef#1{\noexpand\hyperref{}{section}{\prefix}{\prefix}}%
\edef\ewrite{\write\equa{{\string#1},\prefix}%
\write\equa{}}\ewrite%
\writedef{#1\leftbracket#1}
\fi
\else%
\ifnum\nosection=0%
\xdef#1{\noexpand\hyperref{}{section}{\prefix\the\nochapter}%
{\prefix\the\nochapter}}%
\edef\ewrite{\write\equa{{\string#1},\prefix\the\nochapter}%
\write\equa{}}\ewrite%
\writedef{#1\leftbracket#1}
\else%
\ifnum\nosubsection=0%
\xdef#1{\noexpand\hyperref{}{section}{\prefix\the\nochapter.%
\the\nosection}{\prefix\the\nochapter.\the\nosection}}%
\writedef{#1\leftbracket#1}
\edef\ewrite{\write\equa{{\string#1},\prefix\the\nochapter%
.\the\nosection}\write\equa{}}\ewrite%
\else%
\xdef#1{\noexpand\hyperref{}{section}%
{\prefix\the\nochapter.\the\nosection.\the\nosubsection}%
{\prefix\the\nochapter.\the\nosection.\the\nosubsection}}%
\writedef{#1\leftbracket#1}
\edef\ewrite{\write\equa{{\string#1},\prefix\the\nochapter.\the\nosection%
.\the\nosubsection}\write\equa{}}\ewrite%
\fi\fi\fi}%

\newwrite\tfile \def\writetoca#1{}
\def\writetoc{\immediate\openout\tfile=\jobname.tab
\def\writetoca##1{{\edef\next{\write\tfile{\noindent ##1 \string\leaderfill%
\noexpand\number\pageno\par}}\next}}}

%
\def\nolabels{\def\wrlabeL##1{}\def\eqlabeL##1{}\def\reflabeL##1{}}
\def\writelabels{\def\wrlabeL##1{\leavevmode\vadjust{\rlap{\smash%
{\line{{\escapechar=` \hfill\rlap{\sevenrm\hskip.03in\string##1}}}}}}}%
\def\eqlabeL##1{{\escapechar-1\rlap{\sevenrm\hskip.05in\string##1}}}%
\def\reflabeL##1{\noexpand\llap{\noexpand\sevenrm\string\string\string##1}}}
\ifdraftmode\writelabels\else\nolabels\fi

\global\newcount\refno \global\refno=1
\newwrite\rfile
\def\ref{[\hyperref{}{reference}{\the\refno}{\the\refno}]\nref}
\def\nref#1{\DefWarn#1%
\xdef#1{[\noexpand\hyperref{}{reference}{\the\refno}{\the\refno}]}%
\writedef{#1\leftbracket#1}%
\ifnum\refno=1\immediate\openout\rfile=\jobname.ref\fi
\chardef\wfile=\rfile\immediate\write\rfile{\noexpand\item{[\noexpand\hyperdef%
\noexpand\hypernoname{reference}{\the\refno}{\the\refno}]\ }%
\reflabeL{#1\hskip.31in}\pctsign}\global\advance\refno by1\findarg}
\def\findarg#1#{\begingroup\obeylines\newlinechar=`\^^M\pass@rg}
{\obeylines\gdef\pass@rg#1{\writ@line\relax #1^^M\hbox{}^^M}%
\gdef\writ@line#1^^M{\expandafter\toks0\expandafter{\striprel@x #1}%
\edef\next{\the\toks0}\ifx\next\em@rk\let\next=\endgroup\else\ifx\next\empty%
\else\immediate\write\wfile{\the\toks0}\fi\let\next=\writ@line\fi\next\relax}}
\def\striprel@x#1{} \def\em@rk{\hbox{}}
\def\lref{\begingroup\obeylines\lr@f}
\def\lr@f#1#2{\DefWarn#1\gdef#1{\let#1=\UNd@FiNeD\ref#1{#2}}\endgroup\unskip}

\def\addref#1{\immediate\write\rfile{\noexpand\item{}#1}} 
\def\listrefs{{}\vfill\supereject\immediate\closeout\rfile\writestoppt
\baselineskip=14pt
\gdef\reference{\iffrancmode  R\'eferences \else References\fi}
\mark{\reference}  
\nochapter=0\def\sectionname{\reference} 
{\centerline{\bf \hyperdef\hypernoname{refer}{refer}{\bf \reference}}}
\nobreak\bigskip\noindent 
\writetoca{\string\hyperref{}{refer}{refer}{\reference}}
{\parindent=20pt%
\frenchspacing\escapechar=` \input \jobname.ref\vfill\eject}\nonfrenchspacing}
\def\startrefs#1{\immediate\openout\rfile=\jobname.ref\refno=#1}
\def\xref{\expandafter\xr@f}\def\xr@f[#1]{#1}
\def\refs#1{\count255=1[\r@fs #1{\hbox{}}]}
\def\r@fs#1{\ifx\UNd@FiNeD#1\message{reflabel \string#1 is undefined.}%
\nref#1{need to supply reference \string#1.}\fi%
\vphantom{\hphantom{#1}}{\let\hyperref=\relax\xdef\next{#1}}%
\ifx\next\em@rk\def\next{}%
\else\ifx\next#1\ifodd\count255\relax\xref#1\count255=0\fi%
\else#1\count255=1\fi\let\next=\r@fs\fi\next}
%
\newwrite\lfile
{\escapechar-1\xdef\pctsign{\string\%}\xdef\leftbracket{\string\{}
\xdef\rightbracket{\string\}}\xdef\numbersign{\string\#}}
\def\writedefs{\immediate\openout\lfile=\jobname.def \def\writedef##1{%
{\let\hyperref=\relax\let\hyperdef=\relax\let\hypernoname=\relax
 \immediate\write\lfile{\string\def\string##1\rightbracket}}}}%
\def\writestop{\def\writestoppt{\immediate\write\lfile{\string\pageno%
\the\pageno\string\startrefs\leftbracket\the\refno\rightbracket%
\string\def\string\secsym\leftbracket\secsym\rightbracket%
\string\secno\the\secno\string\meqno\the\meqno}\immediate\closeout\lfile}}
\def\writestoppt{}\def\writedef#1{}
\writedefs
\def\biblio\par{\vskip0pt plus.1\vsize\penalty-100\vskip0pt plus-.1
\vsize\bigskip\vskip\parskip
\message{Bibliographie}
{\leftline{\bf \hyperdef\hypernoname{biblio}{bib}{Bibliographical Notes}}}
\nobreak\medskip\noindent\frenchspacing
\writetoca{\string\hyperref{}{biblio}{bib}{Bibliographical Notes}}}%

\def\biblionote{\iffrancmode Notes Bibliographiques\else Bibliographical Notes
\fi}
\def\beginbib\par{\vskip0pt plus.1\vsize\penalty-100\vskip0pt plus-.1
\vsize\bigskip\vskip\parskip
\message{Bibliographie}
{\leftline{\bf \hyperdef\hypernoname{biblio}{\prefix\the\nochapter}%
{\biblionote}}}
\nobreak\medskip\noindent\frenchspacing
\writetoca{\string\hyperref{}{biblio}{\prefix\the\nochapter}%
{\biblionote}}}%

\def\Exercises{\iffrancmode Exercices\else Exercises
\fi}
\def\exerc\par{\vskip0pt plus.1\vsize\penalty-100\vskip0pt plus-.1
\vsize\bigskip\vskip\parskip\global\noexerc=0
\message{Exercises}
\iffrancmode\mark{Exercices}\else\mark{Exercises}\fi
{\leftline{\bf\hyperdef\hypernoname{exercise}{\the\nochapter}{\Exercises}}}
\nobreak\medskip\noindent\frenchspacing
\writetoca{\string\hyperref{}{exercise}{\the\nochapter}{\Exercises}}
}
\def\esubsec{\ifnum\noexerc=0\vskip-12pt\else\vskip0pt plus.05\vsize%
\penalty-100\vskip0pt plus-.05\vsize\bigskip\vskip\parskip\fi%
\global\advance\noexerc by 1
\hyperdef\hypernoname{exercise}{\the\nochapter.\the\noexerc}{}%
\vbox{\noindent\it \iffrancmode Exercice\else Exercise\fi\ \the\nochapter.\the\noexerc}\smallskip\noindent}
\def\exelbl#1{\ifdraftmode{\hfill\escapechar-1{\rlap{\hskip-1mm%
\sevenrm\string#1}}}\fi%
{\xdef#1{\noexpand\hyperref{}{exercise}{\the\nochapter.\the\noexerc}%
{\the\nochapter.\the\noexerc}}}%
\edef\ewrite{\write\equa{{\string#1}\the\nochapter.\the\noexerc}%
\write\equa{}}\ewrite%
\writedef{#1\leftbracket#1}}

\def\eqnn{\global\advance\neqno by 1 \ifinner\relax\else%
\eqno\fi(\prefix\the\nochapter.\the\neqno)}
%
\def\eqnd#1{\DefWarn#1%
\global\advance\neqno by 1 
{\xdef#1{($\noexpand\hyperref{}{equation}{\prefix\the\nochapter.\the\neqno}%
{\prefix\the\nochapter.\the\neqno}$)}}
\ifinner\relax\else\eqno\fi(\hyperdef\hypernoname{equation}{\prefix\the%
\nochapter.\the\neqno}{\prefix\the\nochapter.\the\neqno})
\writedef{#1\leftbracket#1}
\ifdraftmode{\escapechar-1{\rlap{\hskip.2mm\sevenrm\string#1}}}\fi
\edef\ewrite{\write\equa{{\string#1},(\prefix\the\nochapter.\the\neqno)
{\noexpand\number\pageno}}\write\equa{}}\ewrite}
%
\def\checkm@de#1#2{\ifmmode{\def\f@rst##1{##1}\hyperdef\hypernoname{equation}%
{#1}{#2}}\else\hyperref{}{equation}{#1}{#2}\fi}
\def\f@rst#1{\c@t#1a\em@ark}\def\c@t#1#2\em@ark{#1}
\def\eqna#1{\DefWarn#1%
\global\advance\neqno by1\ifdraftmode{\hfill%
\escapechar-1{\rlap{\sevenrm\string#1}}}\fi%
\xdef #1##1{(\noexpand\relax\noexpand%
\checkm@de{\prefix\the\nochapter.\the\neqno\noexpand\f@rst{##1}1}%
{\hbox{$\prefix\the\nochapter.\the\neqno##1$}})}
\writedef{#1\numbersign1\leftbracket#1{\numbersign1}}%
} 
\def\em@rk{\hbox{}} 
\def\xeqn{\expandafter\xe@n}\def\xe@n(#1){#1}
\def\xeqna#1{\expandafter\xe@na#1}\def\xe@na\hbox#1{\xe@nap #1}
\def\xe@nap$(#1)${\hbox{$#1$}}
\def\eqns#1{(\e@ns #1{\hbox{}})}
\def\e@ns#1{\ifx\UNd@FiNeD#1\message{eqnlabel \string#1 is undefined.}%
\xdef#1{(?.?)}\fi{\let\hyperref=\relax\xdef\next{#1}}%
\ifx\next\em@rk\def\next{}%
\else\ifx\next#1\xeqn#1\else\def\n@xt{#1}\ifx\n@xt\next#1\else\xeqna#1\fi
\fi\let\next=\e@ns\fi\next}
\def\figure#1#2{\global\advance\nofigure by 1 \vglue#1%
\hyperdef\hypernoname{figure}{\the\nofigure}{}%
{\elevenpoint
\setbox1=\hbox{#2}
\ifdim\wd1=0pt\centerline{Fig.\ \the\nofigure\hskip0.5mm}%
\else\def\caption{Fig.\ \the\nofigure\quad#2\hskip0mm}
\setbox0=\hbox{\caption}
\ifdim\wd0>\hsize\noindent Fig.\ \the\nofigure\quad#2\else
                 \centerline{\caption}\fi\fi}\par}
\def\lfigure#1#2{\global\advance\nofigure by
1\vglue#1%
\hyperdef\hypernoname{figure}{\the\nofigure}{}%
\leftline{\elevenpoint\hskip10truemm  Fig.\
\the\nofigure\quad #2}} 
\def\figlbl#1{\ifdraftmode{\hfill\escapechar-1{\rlap{\hskip-1mm%
\sevenrm\string#1}}}\fi%
{\xdef#1{\noexpand\hyperref{}{figure}{\the\nofigure}%
{\the\nofigure}}}%
\edef\ewrite{\write\equa{{\string#1}\the\nofigure}%
\write\equa{}}\ewrite%
\writedef{#1\leftbracket#1}}
\def\tablbl#1{\global\advance\notable by
1\ifdraftmode{\hfill\escapechar-1{\rlap{\hskip-1mm%
\sevenrm\string#1}}}\fi%
\hyperdef\hypernoname{table}{\the\notable}{}
{\xdef#1{\noexpand\hyperref{}{table}{\the\notable}%
{\the\notable}}}%
\edef\ewrite{\write\equa{{\string#1}\the\notable}%
\write\equa{}}\ewrite%
\writedef{#1\leftbracket#1}}

\catcode`@=12

\input epsf
\catcode`\À=\active\defÀ{\ifmmode\grave A\else\`A\fi}
\catcode`\É=\active\defÉ{\ifmmode\acute E\else\'E\fi}
\catcode`\à=\active\defà{\ifmmode\grave a\else\`a\fi}
\catcode`\â=\active\defâ{\ifmmode\hat a\else\^a\fi}
\catcode`\ä=\active\defä{\ifmmode\ddot a\else\"a\fi}
\catcode`\ç=\active\defç{\c c}
\catcode`\é=\active\defé{\ifmmode\acute e\else\'e\fi}
\catcode`\è=\active\defè{\ifmmode\grave e\else\`e\fi}
\catcode`\ê=\active\defê{\ifmmode\hat e\else\^e\fi}
\catcode`\ë=\active\defë{\ifmmode\ddot e\else\"e\fi}
\catcode`\ï=\active\defï{\ifmmode\ddot i\else\"\i\fi}
\catcode`\î=\active\defî{\ifmmode\hat i\else\^\i\fi}
\catcode`\ö=\active\defö{\ifmmode\ddot o\else\"o\fi}
\catcode`\ô=\active\defô{\ifmmode\hat o\else\^o\fi}
\catcode`\ù=\active\defù{\ifmmode\grave u\else\`u\fi}
\catcode`\ü=\active\defü{\ifmmode\ddot u\else\"u\fi}
\catcode`\û=\active\defû{\ifmmode\hat u\else\^u\fi}

\input colordvi

\def\g{g}
\def\Abbar{\mkern0.5mu\raise0.5pt\hbox{$\not$}\mkern-0.8mu{\bf A}\mkern.1mu}
\def\Ds{\mkern-0.5mu\raise0.5pt\hbox{$\not$}\mkern-.6mu {\bf D}\mkern.1mu}

\def\*{\par {\bf ************* question/remark/changes made here  ***********}\par }



 \centerline {\bf   A Talk on 3d Matter Coupled to Chern-Simons Field}
\centerline {\bf Spontaneous Breaking of Scale Invariance}
\centerline {\bf and Fermion-Boson Mapping ${}^\dagger$}

\vskip .8cm
\centerline{\bf Moshe Moshe  ${}^{a}$}
\vskip .5cm
\centerline{\bf Technion~-~Israel Inst. of Technology~~, Haifa, Israel} \vskip .8cm

\footnote{} {$\dagger $ Talk presented at the 2019 Meeting of the Division of Particles and Fields of the American Physical Society (DPF2019), July 29 - August 2, 2019, Northeastern University, Boston, C1907293 }
\footnote{} {(a) moshe@technion.ac.il}

\centerline{\bf Abstract} \vskip .8cm

\noindent The singlet sector of vector, large $N$, 3d field theory
corresponds  to Vasiliev higher spin theory on $AdS_4$. Will discuss three dimensional  $U(N)$  symmetric field theory with fermion and boson  matter coupled  to a topological Chern-Simons field. In the presence of a marginal deformation will determine the conditions for the existence of a phase with spontaneous breaking of scale invariance.
In this phase the ground state contains massive $U(N)$ quanta and a massless $U(N)$ singlet bound state goldstone boson:~the Dilaton.
Will show that such a phase appears only in the presence of a marginal deformation.
The massless Dilaton appears in the spectrum provided certain relations between coupling constants are satisfied.  Will discuss the fermion-boson mapping and show that the conditions for spontaneous breaking of scale invariance in the boson and fermion theories are copies of each other.

\vfill\eject



This is a short summary of a talk on spontaneous breaking of scale invariance in three dimensional  $U(N)$  symmetric field theory with fermion and boson  matter coupled  to a topological Chern-Simons field

\section ~~~~~~~~~~~~Motivation

The interest in the phase structure of large $N$  ~$O(N)$  and $U(N)$
symmetric theories with matter fields in the fundamental representation
is focused on the conjectured
  $AdS/CFT$ correspondence  between the singlet
sector of the $O(N)$
vector theory in d=3 space-time dimensions and Vasiliev's higher spin gravity theory
on $AdS_4$.
The quantum completion of the gravitational side of this duality is not known.
If only tree level is considered, one has
 in this case an $AdS/CFT$ correspondence for
which both sides are weakly coupled. The advantage, in this case,
is obvious as a testing ground
of $AdS/CFT$ ideas using explicit derivations.

Interacting scalars and fermions in $O(N)$ and  $U(N)$ symmetric
field theories at large $N$ are well understood  and their phase
structure is known. In the presence of the
Chern-Simons gauge field  the singlet sector is singled out and
the system fits well into the $AdS/CFT$ conjectured duality. Using
large N methods and the convenience of the light-cone gauge for
the Chern-Simons action, explicit calculations can be performed
shedding light on the $AdS/CFT$ correspondence. The $d=3$
$O(N)$ model $g(\vec\phi)^4$ was deformed
 to include a marginal
interaction term $(\lambda_6/24\pi )(\vec\phi)^6$. It was
conjectured that the gravity dual of these 3d theories on the
boundary  is a parity broken version of Vasiliev's higher spin
theory on $AdS_4$ in the bulk with a parity
breaking parameter $\theta$ which depends on the 't Hooft coupling
$\lambda= N/\kappa$.

The theory of scalars and fermions in the fundamental representation of $U(N)$ coupled
to Chern-Simons gauge fields
are dual to the same bulk gravity theory on $AdS_4$. Thus it has also
been conjectured that there is a duality
between the boson and fermion theories. Calculation of
the thermal free energy in large N Chern-Simons field coupled to fermions and scalars
further strengthened this duality.

It was noted however that in the case of massive fermion and scalars the
thermal free energy can be also calculated and the duality between the two Chern-Simons matter
 theories of fermion and boson
may still exist. In this case  high spin symmetry with the bulk Vasiliev's
$AdS_4$ is lost since the boundary theory is not conformal.

It was also noted that the introduction of masses through
spontaneous breaking of scale invariance is an alternative
at large N. Though at finite $N$ the breakdown of the scale symmetry
is explicit it is of ${\cal O}(N^{-1})$, namely the same order by which the $AdS/CFT$ correspondence
is approximated in the above mentioned conjecture.

It is therefore appealing to further analyse the Chern-Simons matter theory
with masses introduced through spontaneous breaking of scale invariance
which assures the introduction of masses into the theory
but leaves the boundary d=3 theory
conformal to ${\cal O}(N^{-1})$.

\section ~~~Conditions for the Spontaneous Breaking of Scale Invariance in Chern-Simons gauge field coupled to a $U(N)$ scalars and fermions

First will consider  a complex scalar field $\phi(x)$ in
the fundamental representation of U(N) in three Euclidean
dimensions coupled to a Chern-Simons level $\kappa$ gauge field
$A_\mu(x)$. To the free Chern-Simons action
$$\eqalignno{{\cal S}_{\rm CS}({\bf A}) =-{i\kappa\over 4\pi}
\epsilon_{\mu\nu\rho}\int\d^3 x\, Tr\left[ {\bf A}_ \mu(x)
\partial_ \nu  {\bf A}_ \rho(x) + \frac{2}{3}  {\bf A}_\mu(x) {\bf
A}_ \nu(x){\bf A}_\rho(x)\right],~~~~&\eqnd\CSfree\cr}$$ \noindent
we add
 the $U(N)$
invariant action
$${\cal S}_{\rm Scalar}=\int\d^3 x\left[  ({\bf D}_\mu{\phi(x)})^\dagger\cdot
{\bf D}_\mu{\phi(x)} +NV({\phi(x)}^\dagger \cdot{\phi(x)} /N) \right],  \eqnd\scalarAction$$
where the complex $\phi(x)$ field is in the fundamental representation and ${\bf D}_\mu$
is the covariant derivative
${\bf D}_\mu=\partial_\mu +{\bf A}_\mu\,.$

The following U(N) invariant self interacting scalar
potential $V({\phi}^\dagger \cdot{\phi} /N)$ will be considered:
$$NV({\phi}^\dagger \cdot{\phi} /N) =
{\mu^2{\phi}^\dagger \cdot{\phi}}
+
{1\over 2}{\lambda_4\over N}({\phi}^\dagger \cdot{\phi})^2+{1\over 6} {\lambda_6 \over N^2}
({\phi}^\dagger \cdot{\phi})^3 \eqnd\Vphi$$
We will concentrate on the
scale invariant potential with renormalized
$\mu_R=\lambda_{4R}=0$ and
analyze the system in its massive phase of spontaneously broken scale invariance.
The CS gauge coupling is defined here as  $\lambda_b=N/\kappa$. A detailed analysis shows that spontaneous breaking of scale invariance occurs when a certain relation is maintain between the $\lambda_6$ coupling of the marginal operator $({\phi}^\dagger \cdot{\phi})^3$ and  the gauge coupling $\lambda_b$


\vskip 1cm

Next will discuss  the  Chern--Simons action coupled to fermion matter in the large $N$ limit.In contrast to previous work that  mainly uses the light-cone gauge, we
  choose the gauge ${\bf A}_3=0$
We add to the Chern--Simons action, quantized in the ${\bf A}_3=0$ gauge, a $U(N)$ gauge-invariant action for an $N$-component spinor field ${\Bfg\psi}$,
$${\cal S}(\psi,\bar\psi,{\bf A})={\cal S}_{\rm CS}({\bf A})-\int\d^3 x\,\bar{\Bfg\psi}(x)(\Ds+M_0){\Bfg\psi}(x)\eqnd{\eCSUNApsi}$$
Here, the CS gauge coupling is defined as
$  {\g\over 4\pi}={N\over \kappa}$

A scalar field $\sigma(x)$ is added  to the action
$${\cal S}_\sigma=\int\d^3 x\left[-\sigma(x)\bar{\Bfg\psi}(x)\cdot{\Bfg\psi}(x)+{N\over 3 g_\sigma}\sigma^3(x)-N {\cal R}\sigma(x)\right],\eqnd\eCSsigmaaction $$
where the new dimensionless parameters $g_\sigma$ and ${\cal R}$ are fixed when $N\to\infty$.\par
The extra $\sigma(x)$  terms in Eq.~\eqns{\eCSsigmaaction} are analogous to the triple-trace deformation $\lambda_6\phi(x)^6$ added to the Chern--Simons boson theory.  The term  ${\cal R}\sigma(x)$ added here in order to have a proper perturbative meaning to the $\sigma$  term in the action.\par
From dimensional analysis, one concludes that $\sigma$ has mass dimension $1$ and that no other term of dimension $3$ or less and odd in $\sigma$ can be added.

In the boson matter  CS theory, when the renormalized parameters $ M_R=\lambda_{4R}=0$, one finds at this scale invariant point a non-zero physical mass,  $  M_{phys}\neq 0 $, as a solution of the gap equation if the following condition is satisfied:

$$\lambda_b^2+{\lambda_6\over 8\pi^2}=4
\,, \eqnd\eCSmassiveBosonic $$
where $\lambda_b$ is the Chern--Simons gauge coupling in the
boson theory, and $\lambda_6$ is the marginal coupling of the
boson $\lambda_6  ({\Bfg \phi}^\dagger\cdot{\Bfg\phi})^3 / 6 N^2$ interaction.

When the forces in this system are tuned according to  Eq.{\eCSmassiveBosonic}, the boson $O(N)$ quanta pairs to create a  massless $O(N)$ singlet excitation. One finds in scattering amplitudes, in the $O(N)$ singlet  channel,  a massless pole, the goldston boson of the spontaneous breaking of scale invariance - the Dilaton. One also finds that the EM trace is $<T_{\mu\mu}>=0$. The vanishing of the EM trace is maintained also at finite temperatures.

In the fermion matter CS system if the renormalized parameters $ M_R=0$, one finds at this scale invariant point a non-zero physical mass  for the fermion quanta, $  M_{phys}\neq 0 $, as a solution of the gap equation, if the following condition is satisfied:
$$\left(\g-g_\sigma \over 2\pi\right)\left(1-{\g \over 8\pi}\right)=1\,. \eqnd\eCSmassiveFermionic  $$
where $g$  here is the Chern--Simons gauge coupling in the
fermion theory, whereas $g_\sigma $ is the marginal coupling in the fermionic case.
Our calculation, in contrast to others, was done in this case in the Weyl Gauge  rather than in the Light Cone gauge.

As in the boson case, when the relation in Eq.{\eCSmassiveFermionic} is maintained, here the fermion $O(N)$ quanta pairs to create a  massless scalar $O(N)$ singlet excitation. One finds in scattering amplitudes, in the $O(N)$ singlet  channel,  a massless pole, the goldston boson of the spontaneous breaking of scale invariance - the Dilaton. Here too, $<T_{\mu\mu}>=0$.

It is now interesting to find out  whether the boson  and fermion  conditions in Eqs.~\eqns{\eCSmassiveBosonic} and \eqns{\eCSmassiveFermionic} are dual copies, when the dual values of the couplings are taken into account.
Indeed, using the fermion-boson dual mapping:

$$\lambda_b={g\over 4\pi}-1    \ , \quad
\lambda_6=8\pi^2\left(1-{g\over4\pi}\right)^2\left(3-4{g\over g_\sigma}\right),
\eqnd\eCSFBmapping
$$	

\noindent one finds that the boson  condition in
Eq.~\eqns{\eCSmassiveBosonic} and the fermion  condition in
Eq.~\eqns{\eCSmassiveFermionic} (derived here in a temporal gauge)
are identical.

.
\vskip.5cm
\noindent The details of this summary can be found in the following papers and references to earlier works :
\vskip.3cm
Bosons and CS:

W. A. Bardeen and  M. Moshe  JHEP 1406 (2014) 113  ~~arXiv:1402.4196 [hep-th]
\vskip.3cm	
Fermions and CS:

M. Moshe and J. Zinn-Justin  JHEP 1501 (2015) 054   ~~arXiv:1410.0558 [hep-th]
\vskip.3cm

\end